\documentclass[aps,twocolumn,showpacs,superscriptaddress]{revtex4}
\usepackage{epsfig}

\begin{document}


\title{Magnetovolume effect in Ce(Ni$_{1-x}$Cu$_x$)$_5$ alloys}

\author{G.E. Grechnev}
\affiliation{B. Verkin Institute for Low Temperature Physics and
Engineering, National Academy of Sciences, 47 Lenin Ave., 61103
Kharkov, Ukraine}

\author{A.V. Logosha}
\affiliation{B. Verkin Institute for Low Temperature Physics and
Engineering, National Academy of Sciences, 47 Lenin Ave., 61103
Kharkov, Ukraine}

\author{A.S. Panfilov}
\email[]{panfilov@ilt.kharkov.ua}
\affiliation{B. Verkin Institute
for Low Temperature Physics and Engineering, National Academy of
Sciences, 47 Lenin Ave., 61103 Kharkov, Ukraine}

\author{I.V. Svechkarev}
\affiliation{B. Verkin Institute for Low Temperature Physics and
Engineering, National Academy of Sciences, 47 Lenin Ave., 61103
Kharkov, Ukraine}

\author{O. Musil}
\affiliation{Charles University, Faculty of Mathematics and
Physics, Department of Electronic Structures, Ke Karlovu 5, 121 16
Prague 2, The Czech Republic}

\author{P. Svoboda}
\affiliation{Charles University, Faculty of Mathematics and
Physics, Department of Electronic Structures, Ke Karlovu 5, 121 16
Prague 2, The Czech Republic}


\begin{abstract}
Magnetic susceptibility $\chi$ of the isostructural
Ce(Ni$_{1-x}$Cu$_x$)$_5$ alloys $(0\le x\le 0.9)$ was studied as a
function of the hydrostatic pressure up to 2 kbar at fixed
temperatures 77.3 and 300~K, using a pendulum-type magnetometer. A
pronounced magnitude of the pressure effect is found to be
negative in sign and strongly (non-monotonously) dependent on the
Cu content, showing a sharp maximum at $x\simeq0.4$. The
experimental results are discussed in terms of the valence
instability of Ce ion in the studied alloys. For the reference
CeNi$_5$ compound the main contributions to $\chi$ and their
volume dependence are calculated ${\it ab~initio}$ within the
local spin density approximation (LSDA), and appeared to be in
close agreement with experimental data.
\end{abstract}


\maketitle

\section{Introduction}

Many of Ce intermetallics are characterized by a strong hybridization of the
magnetic $4f$ electrons with the conduction electron states resulted in
delocalization of the $4f$ level and a change of its occupancy, and hence
the Ce valence. As is evident from measurements of X-ray absorption and
lattice parameters \cite{Gignoux}, together with the magnetic
\cite{Musil,Brandt}, electric and thermoelectric properties \cite{Brandt} in
the isostructural Ce(Ni$_{1-x}$Cu$_x$)$_5$ alloys, the Ce valence decreases
steadily from Ce$^{4+}$ to Ce$^{3+}$ with increase of the Cu content, and
the system undergoes a series of transitions from the nonmagnetic metal with
empty $4f$ level ($x=0$) through the intermediate valence (IV) state
combined with a nonmagnetic dense Kondo state ($0.1\le x\le0.8$) to the
magnetic $4f$ metal ($0.9\le x\le1$). Thus, the reference CeNi$_5$ compound
is the exchange-enhanced itinerant paramagnet
\cite{Gignoux,Coldea,Nordstrom} with the temperature dependent magnetic
susceptibility exhibiting a broad maximum around 100 K, similar to those
observed in YNi$_5$ and LuNi$_5$ \cite{Coldea,Burzo}. On the other side, the
CeCu$_5$ compound is a Kondo lattice antiferromagnet with $T_{\rm N}=3.9$ K
and $T_{\rm K}=2.2$~K \cite{Bauer}. The magnetic susceptibility in CeCu$_5$
at $T\ge 50$ K obeys a Curie-Weiss law with the effective magnetic moment
value close to that expected for Ce$^{3+}$ state \cite{Bauer,Pop,Pop2}. Due
to a direct relation between magnetic properties and the rare earth (RE)
valence state, and also the strong correlation between the valence itself
and RE ionic volume, the RE compounds with unstable $f$ shell exhibit a
large magnetovolume effect. Therefore, a study of pressure effect on
magnetic properties of the systems with variable RE valence is of great
interest to gain insight into a nature of the IV state.

Here we report results of our investigation of the pressure effect on the
magnetic susceptibility of Ce(Ni$_{1-x}$Cu$_x$)$_5$ alloys in a
wide range of Cu concentrations. The experimental results are
supplemented by calculations of the magnetovolume effect value for
the reference CeNi$_5$ compound, using a modified relativistic
full potential approach within linearized "muffin-tin" orbital
method (FP-LMTO).

\section{Experimental details and results}

The polycrystalline samples of Ce(Ni$_{1-x}$Cu$_x$)$_5$ alloys $(0\le x\le
0.9)$ were prepared by arc-melting of a stoichiometric amount of initial
elements in a water cooled crucible under protective argon atmosphere. The
study of X-ray powder diffraction at room temperature revealed that all
samples crystallize in CaCu$_5$-type hexagonal structure, and obtained data
on their lattice parameters agree closely with that published in literature.
Any other phases were not detected within the resolution of the X-ray
technique.

\begin{table}[b]
\caption{The magnetic susceptibilities and their pressure derivative for
Ce(Ni$_{1-x}$Cu$_x$)$_5$ alloys at 77.3 and 300 K.} \label{tab1}
\begin{tabular}{ccccc}
\hline
 x & \multicolumn{2}{c}{$\chi, 10^{-3}$ emu/mol}
&\multicolumn{2}{c}{${\rm d\,ln}\chi/{\rm d\,ln}P, {\rm Mbar}^{-1}$} \\
\cline{2-5}
    & $T=77.3$~K & $T=300$~K & $T=77.3$~K & $T=300$~K \\
\hline 0.0   & 3.29 &  2.12&  $-2.72\pm0.3$ & $-1.93\pm0.3$ \\
 0.1  & 2.74 &  1.47&  $-3.41\pm0.4$ & $-3.02\pm0.4$ \\
 0.2  & 1.55 &  1.09&  $-4.55\pm0.4$ & $-4.93\pm0.3$ \\
 0.3  & 1.11 &  1.08&  $-13.0\pm0.5$ & $-9.93\pm0.5$ \\
 0.4  & 1.47 &  1.26&  $-17.1\pm 1 $ & $-9.5\pm0.5$  \\
 0.5  & 3.67 &  1.87&  $-11.8\pm0.5$ & $-5.52\pm0.5$ \\
 0.6  & 7.85 &  2.47&  $-6.63\pm0.5$ & $-3.28\pm0.3$ \\
 0.7  & 9.55 &  2.78&  $-3.8\pm0.3$  & $-2.03\pm0.3$ \\
 0.9  & 9.93 &  2.76&  $-1.42\pm0.2$ & $-1.26\pm0.2$ \\
\hline
\end{tabular}
\end{table}

The pressure effect on the magnetic susceptibility $\chi$ was measured under
helium gas pressure up to 2 kbar at two fixed temperatures, 77.3 and 300~K,
using a pendulum magnetometer placed into the nonmagnetic pressure cell
\cite{Panfilov}. The relative errors of our measurements, performed in the
magnetic field $H=1.7$ T, did not exceed $0.05\%$.

In Fig.~1 the typical pressure dependencies of the magnetic
susceptibility for Ce(Ni$_{1-x}$Cu$_x$)$_5$ alloys demonstrate a
magnitude of the pressure effect ant its linear behavior.
For each temperature the values of $\chi$ at ambient pressure
and their pressure derivatives, d\,ln$\chi$/d$P$, are listed in
Table~I. Both values include corrections for a weak field
dependence of $\chi$ caused by ferromagnetic impurities, which are
less than 5\%. The negative sign of the pressure effect is consistent
with anticipation that high pressure has to increase the valence,
since the Ce ion in the higher valence (less magnetic) state has a
smaller volume.

\begin{figure}[t]
\includegraphics[width=0.75\linewidth]{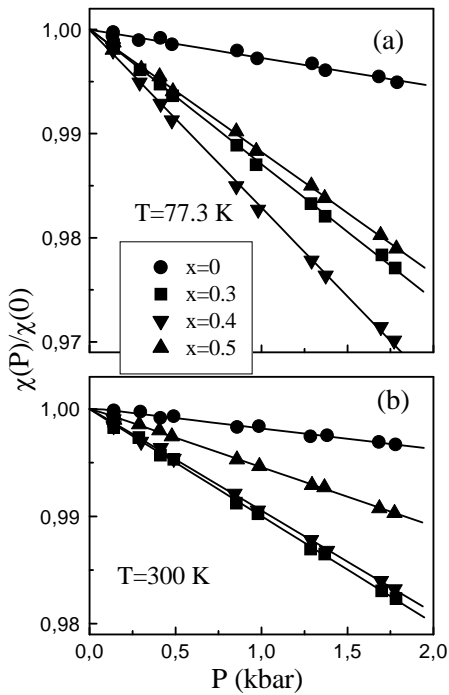}
\caption{Pressure dependence of the magnetic susceptibility of
Ce(Ni$_{1-x}$Cu$_x$)$_5$ alloys at T=77.3~K (a) and 300~K (b)
normalized to its value at $P=0$.} \label{fig1}
\end{figure}

Of particular interest is a strong and non-monotonous concentration
dependence of the pressure effect which shows a sharp maximum in vicinity of
$x\simeq 0.4$ for both temperatures, 77.3 and 300~K (Fig.~2(a)). A
comparison between the obtained experimental results and the data on
concentration dependence of the lattice parameter $a$ and the effective Ce
valence $\nu$ from Ref.\ \onlinecite{Gignoux} (Fig.~2(b)) indicates that the
maximum in d\,ln$\chi(x,T)$/d$P$ correlates with a drastic change of $a$
(and $\nu$) around $x\simeq0.4$ ($\nu\simeq3.5$).
\begin{figure}[ht]
\includegraphics[width=0.75\linewidth]{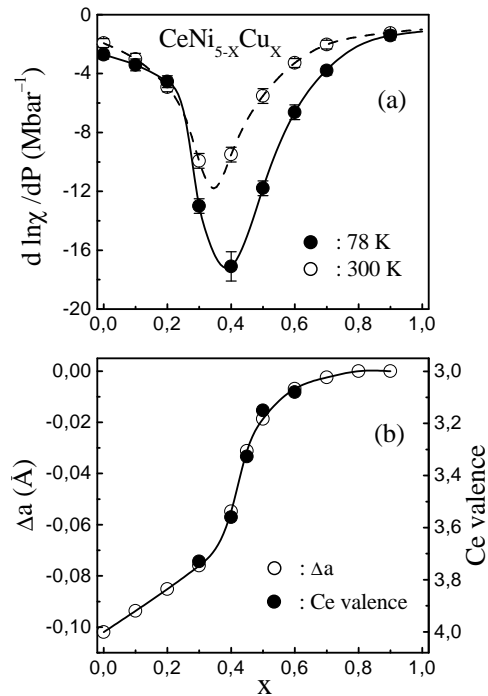}
\caption{(a) Pressure derivative of the magnetic susceptibility
d\,ln$\chi$/d$P$ in Ce(Ni$_{1-x}$Cu$_x$)$_5$ alloys at 77.3 and
300~K. (b) Deviation of the $a$ lattice parameter, $\Delta a$, in
Ce(Ni$_{1-x}$Cu$_x$)$_5$ alloys from the $a(x)$ dependence for the
Ce ion assumed to be in a trivalent state (left scale) and the Ce
valence deduced from X-ray absorption studies (right scale) at
room temperature versus Cu content $x$ (according to the data of
Ref.\protect\onlinecite{Gignoux}).} \label{fig2}
\end{figure}

It is interesting to note that a similar peculiarity in
d\,ln$\chi$/d$P$ versus valence was observed for various Yb
compounds at room temperature \cite{Zell}. As was shown, the
relative change of $\chi$ with pressure is the most pronounced
also at the half-integer value of valence, $\nu\simeq 2.5$, but
contrary to the Ce compounds, it has a positive sign, as can be
expected.

\section{Theory for CeNi$_5$}

$Ab~initio$ calculations of the electronic structure were carried out for
the reference compound CeNi$_5$ by employing a modified FP-LMTO method
\cite{Wills,Buschow}. The exchange-correlation potential was treated in the
LSDA approximation \cite{Barth} of the density functional theory. To analyze
the observed magnetovolume effect value in CeNi$_5$, the magnetic
susceptibility and its volume dependence were calculated within the modified
method, wherein the external magnetic field $H$ was taken into account by
means of the Zeeman operator, $H$(2\^s+\^l). The latter was incorporated in
FP-LMTO Hamiltonian \cite{Grechnev} for calculations of the field-induced
spin and orbital magnetic moments. The corresponding contributions to
magnetic susceptibility were derived from the field-induced moments, which
have been calculated in an external magnetic field of 10~T.

The electronic structure calculations were performed for a number of
lattice parameters close to the experimental one (the ratio $c/a$ was
fixed at its experimental value).
The equilibrium lattice spacing $a_{\rm th}$ and corresponding theoretical
bulk modulus $B_{\rm th}$ were determined from dependence of the total
energy on the unit cell volume, $E(V)$, by using the well known
Murnaghan equation, and appeared to be $a_{\rm th}=8.96$~a.u. and
$B_{\rm th}=1.9$~Mbar.
The Murnaghan equation is based on the assumption that the pressure
derivative $B'$ of the bulk modulus $B$ is constant.
By using the evaluated from the Murnaghan equation value of $B'=3.73$,
we have estimated $B_{\rm th}^{\rm est}=1.45$, corresponding to the
experimental $a_{\rm exp}=9.2$ a.u. \cite{Gignoux}.
This estimation appeared to be in nice agreement with the available
experimental value, $B_{\rm exp}=1.43$~Mbar \cite{Butler}.
The differences between the equilibrium theoretical $a_{\rm th}$ and
$B_{\rm th}$ and experiment on bulk properties of CeNi$_5$ are presumably
related to the overbonding tendency of the LSDA approach \cite{Wills}.

The strongly volume dependent spin contribution to $\chi$ originates
predominantly from the $3d$ states of Ni. Regarding the orbital contribution
to $\chi$, it comes mainly from electrons in the atomic sphere of Ce and
amounts to about 20\% of total susceptibility. At the theoretical lattice
parameter, the calculated total susceptibility ($2.9\times10^{-3}$ emu/mole)
appeared to be very close to the experimental value ($3.0\times10^{-3}$
emu/mole at $T=0$~K \cite{Musil,Coldea}). The calculated volume derivative
of susceptibility, d\,ln$\chi$/d\,ln$V$ = 4.2, is in agreement with that
resulted from the experimentally observed pressure derivative for CeNi$_5$
at $T=77.3$~K, d\,ln$\chi$/d\,ln$V=3.9\pm0.4$. Thus it has been
demonstrated, that LSDA provides an adequate description of the strongly
exchange enhanced magnetic susceptibility of CeNi$_5$ and its pressure
dependence.

\section{Discussion}

As is shown, the LSDA allows to describe the magnetovolume effect
in the reference CeNi$_5$ compound that gives grounds for future
application of $ab~initio$ LSDA approaches to some
Ce(Ni$_{1-x}$Cu$_x$)$_5$ alloys. Here, however, we shall restrict
our consideration of the experimental data in alloys within a
phenomenological approach.

\subsection{Concentration dependence}

Anticipating the pressure effect on the magnetic susceptibility to arise
mainly from the change of Ce valence $\nu$, or the fractional occupation of
the $4f^1$ magnetic state $n_{\rm 4f}$ ($\nu=4-n_{\rm 4f}$), the pressure
effect can be analyzed within a simple relation
\begin{eqnarray}
{{\rm d\,ln}\chi(T)\over{\rm d}P}\approx {\partial\,{\rm
ln}\chi(T)\over\partial n_{\rm 4f}}\times {{\rm d}n_{\rm
4f}\over{\rm d}P} \label{dn_4f/dP}
\end{eqnarray}
in terms of the pressure dependence of $n_{\rm 4f}$ (or $\nu$).
The most reliable results of such analysis would be expected in
the Cu-rich alloys at low temperatures where the $4f$ contribution
$\chi_{\rm 4f}$ becomes dominant ($\chi\approx\chi_{\rm 4f}$). In
Fig.~3(a) the $\chi$ versus $n_{\rm 4f}$ dependence is shown for
Ce(Ni$_{1-x}$Cu$_x$)$_5$ alloys ($0.4\le x\le1$) at 77.3~K, which
was obtained by using the experimental $\chi(x)$ values from Table~1
and $\nu(x)$ data of Fig.~2(b). A substitution of the resulted from
Fig.~3(a) derivatives $\partial\,{\rm ln}\chi/\partial n_{\rm 4f}$
and experimental data on d\,ln$\chi$/d$P$ at 77.3~K into
Eq.~(\ref{dn_4f/dP}) gives the value d$n_{\rm 4f}$/d$P$ which
strongly depends on $\nu_{\rm 4f}$ (Fig.~3(b)). As is seen, the
maximum value of d$n_{\rm 4f}$ is expected at $\nu_{\rm 4f}=0.5$
($\nu=3.5$) to be about $-6.5\pm1.5$ Mbar$^{-1}$. The
corresponding estimates of the valence change under pressure,
d$\nu$/d$P=-$d$n_{\rm 4f}$/d$P$, are of the same order that those
resulted from the study of magnetovolume effect in SmB$_6$
(2~Mbar$^{-1}$ \cite{Panfilov2}) and from the measurements of
resonant inelastic X-ray emission in YbAl$_2$ under pressure
($\sim 5$ Mbar$^{-1}$ \cite{Dallera}).

\begin{figure}[t]
\includegraphics[width=0.75\linewidth]{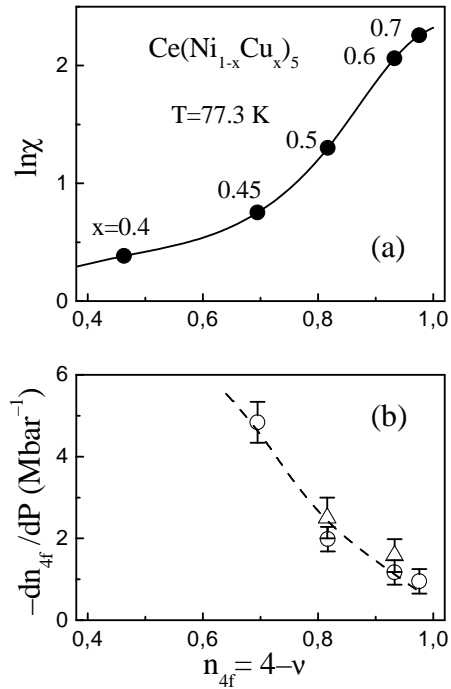}
\caption{(a) Values of ln$\chi$ at 77.3~K. (b) d$n_{\rm 4f}$/d$P$ plotted
against $n_{\rm 4f}$ for Ce(Ni$_{1-x}$Cu$_x$)$_5$ alloys,
\protect{\large$\circ$} and $\triangle$ denote the data obtained within
Eq.~(\protect\ref{dn_4f/dP}) and Eq.~(\protect\ref{dlnX}), respectively.
Points for $x=0.45$ are interpolation of the experimental data on
concentration dependence of $\chi$ and d\,ln$\chi$/d$P$.} \label{fig3}
\end{figure}

\subsection{Temperature dependence}

In a simple empirical model which includes interconfiguration
fluctuations between $f^{n+1}$ and $f^n$ levels \cite{Sales}, the
contribution of the $4f^0$ (J=0) and $4f^1$ (J=5/2) states of Ce
to magnetic susceptibility is given by
\begin{eqnarray}
\chi_{\rm 4f}(T)=N_{\rm A}\mu^2n_{\rm 4f}(T)/3k(T+T_{\rm f}).
\label{T_f}
\end{eqnarray}
Here $N$ is the Avogadro number, $\mu$ effective magnetic moment of the
$4f^1$ state, $T_{\rm f}$ the characteristic temperature (valence
fluctuation temperature, or Kondo temperature, or heavy-fermion bandwidth).
It should be mentioned that a quantitative analysis of the $\chi_{\rm
4f}(T)$ dependence using Eq.~(\ref{T_f}) requires the complete data on
$n_{\rm 4f}(T)$ (and probably on $T_{\rm f}(T)$ as well) which are actually
unavailable. Furthermore, to separate the $\chi_{\rm 4f}(T)$ term from the
experimental data on $\chi(T)$ one needs to know a background contribution
$\chi_0$, which generally can not be neglected. A simplified analysis of the
experimental data can be performed assuming $n_{\rm 4f}$, $T_{\rm f}$ and
$\chi_0$  to be temperature independent. Then the magnetic susceptibility
obeys a modified Curie-Weiss law,
\begin{eqnarray}
\chi(T)=\chi_0+\chi_{\rm 4f}(T)\equiv\chi_0+{C\over(T-\Theta)}~,
\label{CW}
\end{eqnarray}
with $C=N\mu^2n_{\rm 4f}/3k$ and $\Theta=-T_{\rm f}$. For the
representative Ce(Ni$_{0.5}$Cu$_{0.5}$)$_5$ alloy, the best fit of
Eq.~(\ref{CW}) to the experimental data at $T\ge 50$ K (Fig.~4(a))
\cite{Musil} is obtained with $\chi_0=0.6\times10^{-3}$ emu/mole,
$C=0.48$ K$\cdot$emu/mole and $\Theta=-79$ K. It should be pointed
out that the estimate $n_{\rm 4f}=0.6$, resulted from $C$, is in a
reasonable agreement with the value of 0.8 that follows from the
data in Fig.~2(b) for $x=0.5$.

As is evident from Eqs.~(\ref{T_f}) and (\ref{CW}), the pressure
effect on the $4f$ susceptibility is governed by changes of
$n_{\rm 4f}$ and $T_{\rm f}$ with pressure, as
\begin{eqnarray}
{{\rm d\,ln}\chi_{\rm 4f}(T)\over{\rm d}P} = {{\rm
d\,ln}C\over{\rm d}P}- {1\over(T+T_{\rm f})}\times{{\rm d}T_{\rm
f}\over{\rm d}P}\equiv\nonumber\\ \equiv{{\rm d\,ln}n_{\rm
4f}\over{\rm d}P}- {\chi_{\rm 4f}(T)\over C}\times{{\rm d}T_{\rm
f}\over{\rm d}P}, \label{dlnX}
\end{eqnarray}
being a linear function of 1/($T+T_{\rm f}$) or $\chi_{\rm
4f}(T)$. The data on d\,ln$\chi_{\rm 4f}$/d$P$ for
Ce(Ni$_{0.5}$Cu$_{0.5}$)$_5$ alloy were derived from the measured
effect, d\,ln$\chi$/d$P$, in the framework of Eq.~(\ref{CW}) by using
a value $B\sim 1$~Mbar$^{-1}$ as a rough estimate for the pressure
dependence of the background \cite{Heine}, which is assumed to
originate from $3d(4d)$ itinerant electrons. The obtained values
are plotted in Fig.~4(b)
\begin{figure}[t]
\includegraphics[width=0.75\linewidth]{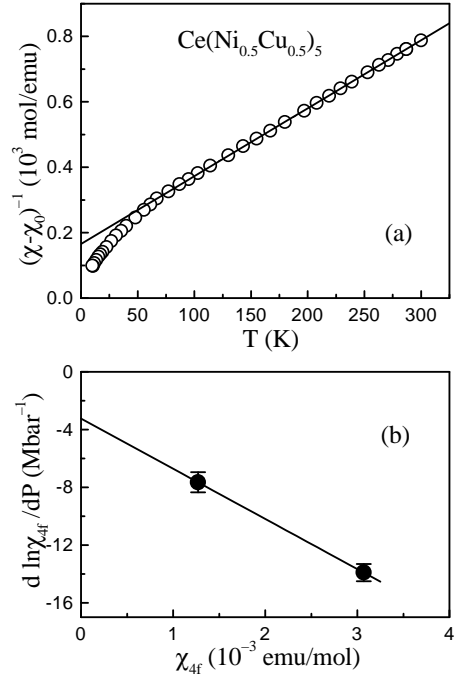}
\caption{Temperature dependence of the magnetic susceptibility
$\chi$ (a) and pressure derivative d\,ln$\chi_{\rm 4f}$/d$P$ plotted
against $\chi_{\rm 4f}$ (b) for Ce(Ni$_{0.5}$Cu$_{0.5}$)$_5$ alloy.}
\label{fig4}
\end{figure}
as a function of $\chi_{\rm 4f}(T)$. Its
linear approximation in accordance with Eq.~(\ref{dlnX}) gives
\begin{eqnarray}
{{\rm d\,ln}C\over{\rm d}P}= {{\rm d\,ln}n_{\rm 4f}\over{\rm d}P}=
-3.2\pm 0.7~{\rm Mbar}^{-1}, \nonumber\\ {{\rm d}T_{\rm f}
\over{\rm d}P}=1650\pm 250~{\rm K\cdot Mbar}^{-1}. \label{dlnT_f}
\end{eqnarray}
The resultant value d$n_{\rm 4f}$/d$P=-2.5\pm0.5$ Mbar$^{-1}$ is
in line with the value d$n_{\rm 4f}=-2.0\pm0.3$ Mbar$^{-1}$
obtained previously for $x=0.5$ from analysis of the concentration
dependence of the pressure effect within Eq.~(\ref{dn_4f/dP}).
From the pressure dependence of $T_{\rm f}$ the corresponding
Gr\"uneisen parameter, $\Omega$, is estimated as
\begin{eqnarray}
\Omega_f\equiv-{{\rm d\,ln}T_{\rm f}\over{\rm d\,ln}V}= B{{\rm
d\,ln}T_{\rm f}\over{\rm d}P}=31\pm5 \label{Omega}
\end{eqnarray}
using the experimental bulk modulus $B=1.5$ Mbar \cite{Staub}. The
Anderson impurity model provides the Kondo temperature and its
pressure derivative to be described in terms of $n_{\rm 4f}$
\cite{Gunnarsson}:
\begin{eqnarray}
T_{\rm K}\propto{1-n_{\rm 4f)}\over n_{\rm 4f}},~~ {{\rm
d\,ln}T_{\rm K}\over {\rm d}P}= {1\over 1-n_{\rm 4f}}\times{{\rm
d\,ln}n_{\rm 4f}\over{\rm d}P}.
\label{T_K}
\end{eqnarray}
Then, assuming $T_{\rm f}\propto T_{\rm K}$ and using in
Eq.~(\ref{T_K}) the values d\,ln$n_{\rm 4f}$/d$P=-3.2\pm0.7$
Mbar$^{-1}$ and $n_{\rm 4f}=0.8$ evaluated above for the alloy
with $x=0.5$, one obtains
\begin{eqnarray}
\Omega_f\simeq \Omega_{\rm K}= -{{\rm d\,ln}T_{\rm K}\over {\rm
d\,ln}V}= 24\pm 5
\end{eqnarray}
in reasonable agreement with the direct estimate (\ref{Omega}).

For Ce(Ni$_{0.4}$Cu$_{0.6}$)$_5$ alloy, the analogous analysis in
the framework of Eq.~(\ref{CW}) and Eq.~(\ref{dlnX}) yields the
following Curie-Weiss parameters: $C\simeq0.806$ K$\cdot$emu/mole,
$\chi_0\sim 0$, $T_{\rm f}=-\Theta=26$ K, and their pressure
derivatives:
\begin{eqnarray*}
{{\rm d\,ln}C\over{\rm d}P}= {{\rm d\,ln}n_{\rm 4f}\over{\rm d}P}=
-1.7\pm 0.5~{\rm Mbar}^{-1},\\ {{\rm d}T_{\rm f}\over{\rm d}P}
=620\pm 100~{\rm K\cdot Mbar}^{-1}~.
\end{eqnarray*}
The latter results in the Gr\"uneisen parameter $\Omega_{\rm
f}=35\pm6$, assuming the bulk modulus value $B=1.5$ Mbar, as in
the Ce(Ni$_{0.5}$Cu$_{0.5}$)$_5$ alloy. The similar estimate
follows from Eqs.~(\ref{T_K}) and (\ref{Omega}) with $n_{\rm
4f}=0.93$ derived from the data in Fig.~2(b). The reasonable
description of the Gr\"uneisen parameter for alloys with $x=0.5$
and 0.6 with Anderson model \cite{Gunnarsson} allows to consider
the Cu-rich alloys studied in the present work as the nonmagnetic
Kondo lattices.

\section{Conclusions}

The pressure effect on magnetic susceptibility of Ce(Ni$_{1-x}$Cu$_x$)$_5$
alloys has been observed for the first time. This effect is negative in
sign, and also strongly and non-monotonously dependent on the Cu content.
For the reference CeNi$_5$ compound, the pressure effect value is
successfully described within LSDA approximation, using the modified full
potential relativistic FP- LMTO method. For Ce(Ni$_{1-x}$Cu$_x$)$_5$ alloys
the effects of pressure and alloying on the valence state of Ce ion are the
most pronounced around $x\sim 0.4$, which corresponds to the half-integer
valence $\nu\sim 3.5$. In other words, the fractional occupation $n_{\rm
4f}\sim 0.5$ with the nearly degenerate $f^0$ and $f^1$ configurations of
electronic states is favorable for the valence instability. It is also found
that the main contributions to the pressure effect on magnetic
susceptibility for the Cu-rich alloys are i) the decrease of the effective
Curie constant and ii) the increase of the characteristic temperature
$T_{\rm f}$. The latter exhibits a large and positive value of the
Gr\"uneisen parameter, which can be apparently described within the Anderson
impurity model. Both of these contributions have their origin in the change
of the Ce valence state caused by depopulation of the $f$ state under
pressure due to its shift relative to the Fermi energy.

\acknowledgments

The work of P.S. and O.M. is a part of the research program MSM
0021620834 financed by the Ministry of Education of the Czech
Republic. The authors thank S.N. Dolya for discussions and V.A.
Desnenko for help in magnetic measurements.



\begin{thebibliography}{00}

\bibitem{Gignoux} D. Gignoux, F. Givord, R. Lemaire, H. Launois,
and F. Sayetat, J. Physique {\bf 43}, 173 (1982).
\bibitem{Musil} O. Musil, P. Svoboda, M. Divi\v{s}, and V. Sechovsk\'y,
Czech. J. Phys. {\bf 51}, Suppl. D, D311 (2005).
\bibitem{Brandt} N.B. Brandt, V.V. Moshchalkov, N.E. Sluchanko, E.M.
Savitskii, and T.M. Shkatova, Solid State Phys. {\bf 26}, 2110 (1984).
\bibitem{Coldea} M. Coldea, D. Andreica, M. Bitu, and V. Crisan,  J. Magn.
Magn. Mater. {\bf 157/158}, 627 (1996).
\bibitem{Nordstrom} L. Nordstr\"om, M.S.S. Brooks, and B. Johansson,
Phys. Rev. B {\bf 46}, 3458 (1992).
\bibitem{Burzo} E. Burzo, V. Pop, and I. Costina, J. Magn. Magn. Mater.
{\bf 157/158}, 615 (1996).
\bibitem{Bauer} E. Bauer, M. Rotter, L. Keller, P. Fisher, M. Ellerby,
and K.M. McEwen, J. Phys.: Condens. Matter {\bf 6}, 5533 (1994).
\bibitem{Pop} I. Pop, R. Pop, and M. Coldea, J. Phys. Chem. Solids
{\bf 43}, 199 (1982).
\bibitem{Pop2} I. Pop, E. Rus, M. Coldea, and O. Pop, J. Phys. Chem.
Solids {\bf 40}, 683 (1979).
\bibitem{Panfilov} A. S. Panfilov, Physics and Technique of High
Pressures (in Russian) {\bf 2}, N2, 61 (1992).
\bibitem{Zell} W. Zell, R. Pott, B. Roden, and D. Wohlleben, Solid State
Commun. {\bf 40}, 751 (1981).
\bibitem{Wills} J.M. Wills, O. Eriksson, ed. by H. Dreysse, Electronic
Structure and Physical Properties of Solids, Springer, Berlin (2000), p. 247.
\bibitem{Buschow} K.H.J. Buschow, G.E. Grechnev, A. Hjelm et al., J. Alloys
and Compounds {\bf 244}, 113 (1996).
\bibitem{Barth} U. von Barth and L. Hedin, J. Phys. C {\bf 5}, 1629 (1972).
\bibitem{Grechnev} G.E. Grechnev, R. Ahuja, and O. Eriksson, Phys. Rev. B
{\bf 68}, 64414 (2003).
\bibitem{Butler} B. Butler, D. Givord, F. Givord, and S.B. Palmer, J. Phys.
C {\bf 13}, L743 (1980).
\bibitem{Panfilov2} A. S. Panfilov, I.V. Svechkarev, Yu. B. Paderno, E.S.
Konovalova, and V.I. Lazorenko, Physics and Technique of High Pressures (in
Russian), issue {\bf 20}, 3 (1985).
\bibitem{Dallera} C. Dallera, E. Annese, J.-P. Rueff, et al., J. Phys.:
Condens. Matter {\bf 17}, S849 (2005).
\bibitem{Sales} B.C. Sales and D. Wohlleben, Phys. Rev. Lett. {\bf 35},
1240 (1975).
\bibitem{Heine} V. Heine, Phys. Rev. {\bf 153}, 673 (1967).
\bibitem{Staub} U. Staub, C. Schulze-Briese, P.A. Alekseev, et al., J.
Phys.: Condens. Matter. {\bf 13}, 11511 (2001).
\bibitem{Gunnarsson} O. Gunnarsson and K. Sch\"onhammer, Phys. Rev. B {\bf 28},
4315 (1983).
\end{thebibliography}
\end{document}